\documentclass[preprint,showpacs]{revtex4}
\usepackage[T2A]{fontenc}
\usepackage[cp1251]{inputenc}
\usepackage[english]{babel}
\usepackage{amsmath}
\usepackage{amsfonts}
\usepackage{graphicx,epsfig}
\usepackage{subfigure}
\usepackage{amssymb}
\usepackage[indent,small]{caption2}
\usepackage{graphicx}
\usepackage{wrapfig}

\begin{document}

\begin{center}
{\Large \textbf{Bose-Einstein Condensation and Ferromagnitism of Low Density
Bose gas of Particles with Arbitrary Spin. }}

{V.S. Babichenko} {I.\,Ya. {Polishchuk}$^{1,2}$}
\end{center}

{$^1$ RRC Kurchatov Institute, Kurchatov Sq., 1, 123182 Moscow, Russia}

{$^2$ Moscow Institute for Physics and Technology, 141700, 9, Institutskii
per., Dolgoprudny, Moscow Region, Russia}

\begin{center}
\today

ABSTRACT
\end{center}

\textit{Properties of the ground state and the spectrum of elementary
excitations are investigated for the low density ultracold spinor 3D Bose
gas of particles with arbitrary nonzero spin. Gross-Pitaevskii equations are
derived. Within the framework of the considering interaction Hamiltonian it
is shown that the ground state spin structure and spin part of the chemical
potential is determined by the renormalized interaction, being defined by
the contribution of the virtual large momenta. The ferromagnetic structure
of the ground state, and the equation of the phase, density, and spin
dynamics are obtained from Gross-Pitaevskii equations.}

\vspace{1cm}

Spinor Bose gas (SBG) at low temperatures is a quantum system in which in
addition to the superfluidity the magnetic order can take place. The study
of such systems gives rise to the great interest from both theoretical and
experimental points of view \cite{2}, \cite{4}, \cite{5}, \cite{Ho}, \cite%
{Mach}, \cite{nuovo}, \cite{PRL113}, \cite{prl116}, \cite{PRL117}, \cite%
{prl119}. In such quantum system a magnetic order and a superfluidity may be
revealed simultaneously. In this paper low density nonideal 3D SBG of
partcles with arbitrary spin $S\neq 0$ is investigated at zero temperature.
The basis of the spin coherent states (SCS) for spin $S$ \cite{Radcl}, \cite%
{Lieb}, \cite{7}\ is used to expand the creation and annihilation operators
of particles $\widehat{\psi }^{+},\widehat{\psi }$ instead of the states
being the eigen states of the spin operator z-projection $\widehat{s}_{z}$.
Hitherto, SCS were employed, as a rule, for investigation of many-body
systems of interacting localized spins \cite{Pol}, \cite{PolWig}, \cite{8},
\cite{9}, \cite{9-2}, \cite{9-1}, \cite{Fulde}, \cite{10}. In present paper
SCS are used for the investigation of Bose system consists of moving
particles with spin $S$. The bare interaction between particles is supposed
to consist of short range and long range parts. The short range bare
interaction depending on the difference between coordinates of two
interacting particles is supposed to have the delta-function dependence on
the coordinates, i.e., the short range interaction is considered as
independet of momentum transfer. The short range interaction is divided into
two parts. The first one is independent of spin interaction with the
coupling constant $g^{\left( 0\right) }$. The second one is the spin
dependent part with the coupling constant $\gamma _{sh}^{\left( 0\right) }$.
The delta-function dependence on the difference between coordinates of two
interacting particles for short range interaction is conditioned by the
smallness of the interaction radius $a_{0}$.The radius $a_{0}$ is supposed
to be the smallest length scale parameter. Properties of low density Bose
gas is considered in the present paper so the length $a_{0}$ obeys the
inequality $a_{0}<<\rho ^{-1/3}$, where $\rho $ is the density of the
system. The radius $a_{0}$ is supposed to be of the same order as for the
spin independent and the spin dependent interactions. The long range part of
the interacton is supposed to be of the dipole-dipole type spin-spn
interacton. The dipole-dipole form of the long range interaction is valid
for suffisiently large distances $\overrightarrow{R}$ between interacting
particles $\mid \overrightarrow{R}\mid =\mid \overrightarrow{r}-%
\overrightarrow{r}^{\prime }\mid >>a_{s}>>a_{0}$, where $\overrightarrow{r},%
\overrightarrow{r}^{\prime }$ are coordinates of interacting particles. The
characteristic length $a_{s}$ is supposed to obey the inequality $%
a_{0}<<a_{s}<<\rho ^{-1/3}$.\ Gross-Pitaevskii equations (GPE) for low
energy degrees of freedom are derived. A spin structure of the Bose-Einstein
condensate (BEC) is found. It is shown that for the model of the interaction
Hamiltonian under consideration the spin structure of the system has a
ferromagnetic order, at that, the renormalization of the bare short range
and long range interactions between long wave Bose fields playes the key
role for the formation of the ground state spin structure. This
renormalization is connected with the contribution of large virtual momenta
during the scattering process. The equation describing the spin density
dynamics is obtained. It is shown that this equation is analogous to Landau
- Lifshits equation for Heisenberg ferromagnet system of the localized
spins. Green functions and spectrums of the elementary exitations of two
types are found.These exitations are the density-phase fluctuations and the
spin fluctuations.

The model of the bare Hamiltonian under consideration is
\begin{equation}
\widehat{H}=\widehat{H}^{\left( 0\right) }+\widehat{H}_{int}^{\left(
0\right) \left( sh\right) }+\widehat{H}_{int}^{\left( 0\right) \left( sp\mid
sh\right) }+\widehat{H}_{int}^{\left( 0\right) \left( l\right) }  \label{H}
\end{equation}

The term $\widehat{H}^{\left( 0\right) }$ is the Hamiltonian of
noninteractng particles

\begin{equation}
\widehat{H}^{\left( 0\right) }=\int d^{3}r\sum\limits_{m=-S}^{S}\widehat{%
\psi }_{m}^{+}\left( t,\overrightarrow{r}\right) \left[ -\frac{1}{2m_{0}}%
\overrightarrow{\nabla }^{2}-\mu \right] \widehat{\psi }_{m}\left( t,%
\overrightarrow{r}\right)  \label{H0}
\end{equation}

where $\mu $ is a chemical potential, $m_{0}$ is a mass of the particle, $%
\widehat{\psi }_{m}$, $\widehat{\psi }_{m}^{+}$ are anihilation and creation
operators, where $m$ is the integer eigenvalues of the z-projection of the
spin operator $\widehat{s}^{\left( z\right) }$, such that $-S\leqslant
m\leqslant S$, and the states $\mid m>$ are the eigenstates of this
operator. The term $\widehat{H}_{int}^{\left( 0\right) \left( sh\right) }$
is the Hamiltonian of the short range independent of spin interaction. This
interection has the form

\begin{equation}
\widehat{H}_{int}^{\left( 0\right) \left( sh\right) }=g^{\left( 0\right)
}\int d^{3}r\sum\limits_{m_{1},m_{2}}\widehat{\psi }_{m_{2}}^{+}\left( t,%
\overrightarrow{r}\right) \widehat{\psi }_{m_{1}}^{+}\left( t,%
\overrightarrow{r}\right) \widehat{\psi }_{m_{1}}\left( t,\overrightarrow{r}%
\right) \widehat{\psi }_{m_{2}}\left( t,\overrightarrow{r}\right)  \label{H1}
\end{equation}

The value $g^{\left( 0\right) }$ is the bare coupling constant of the
independent of spin short range interaction. The radius of the short range
interaction is of the order of the length $a_{0}$, such that $a_{0}<<\rho
^{-1/3}$ due to the smallness of the density $\rho $ of the system being
considered. Moreover, the coupling constants $g^{\left( 0\right) }$ is
supposed to be positive and sufficiently large, so that it obeys the
enequlity $\frac{g^{\left( 0\right) }m_{0}}{a_{0}}>>1$. Note, that the
inequality $\frac{g^{\left( 0\right) }m_{0}}{a_{0}}>>1$ means that the bare
coupling constant $g^{\left( 0\right) }$ should be much larger than the
renormalized spin independent scattering amplitude being of the order of $%
a_{0}/m_{0}$.

The term $\widehat{H}_{int}^{\left( 0\right) \left( sp\mid sh\right) }$ is
the Hamiltonian of the short range spin dependent interaction. This part of
the Hamiltonian is taken in the form

\begin{equation}
\widehat{H}_{int}^{\left( 0\right) \left( sp\mid sh\right) }=\gamma
_{sh}^{\left( 0\right) }\int d^{3}r\sum\limits_{i=1}^{3}\sum\limits
_{\substack{ m_{1},m_{2},  \\ m_{1}^{\prime },m_{2}^{\prime }}}\widehat{s}%
_{m_{2,}m_{2}^{\prime }}^{\left( i\right) }\widehat{s}_{m_{1,}m_{1}^{\prime
}}^{\left( i\right) }\widehat{\psi }_{m_{1}}^{+}\left( t,\overrightarrow{r}%
\right) \widehat{\psi }_{m_{2}}^{+}\left( t,\overrightarrow{r}\right)
\widehat{\psi }_{m_{2}^{\prime }}\left( t,\overrightarrow{r}\right) \widehat{%
\psi }_{m_{1}^{\prime }}\left( t,\overrightarrow{r}\right)  \label{H2}
\end{equation}

The radius of the short range spin dependent interaction is supposed to be
of the order of $a_{0}$. The term $\widehat{H}_{int}^{\left( 0\right) \left(
l\right) }$ is the Hamiltonian of the long range spin-spin interaction of
the dipole-dipole type

\begin{equation}
\widehat{H}_{int}^{\left( 0\right) \left( l\right) }=\int
d^{3}rd^{3}r^{\prime }\sum\limits_{\substack{ m_{1},m_{2},  \\ m_{1}^{\prime
},m_{2}^{\prime }}}\widehat{s}_{m_{1,}m_{1}^{\prime }}^{\left( i\right) }%
\widehat{s}_{m_{2,}m_{2}^{\prime }}^{\left( j\right) }V_{ij}^{\left(
0\right) \left( l\right) }\left( \overrightarrow{R}\right) \widehat{\psi }%
_{m_{1}}^{+}\left( t,\overrightarrow{r}\right) \widehat{\psi }%
_{m_{2}}^{+}\left( t,\overrightarrow{r}^{\prime }\right) \widehat{\psi }%
_{m_{2}^{\prime }}\left( t,\overrightarrow{r}^{\prime }\right) \widehat{\psi
}_{m_{1}^{\prime }}\left( t,\overrightarrow{r}\right)  \label{Hlong}
\end{equation}

where $V_{ij}^{\left( 0\right) \left( l\right) }\left( \overrightarrow{R}%
\right) =\gamma _{l}^{\left( 0\right) }\frac{\mid \overrightarrow{R}\mid
^{2}\delta _{i,j}-3R^{\left( i\right) }R^{\left( j\right) }}{\mid
\overrightarrow{R}\mid ^{5}}=-\gamma _{l}^{\left( 0\right) }\nabla _{%
\overrightarrow{r}}^{\left( i\right) }\nabla _{\overrightarrow{r}}^{\left(
j\right) }\frac{1}{\mid \overrightarrow{R}\mid },$ $\gamma _{l}^{\left(
0\right) }>0$, $\overrightarrow{R}=\overrightarrow{r}-\overrightarrow{r}%
^{\prime }$, and $\overrightarrow{r},\overrightarrow{r}^{\prime }$ are the
space coordinates of the interacting particles, $\nabla _{\overrightarrow{r}%
}^{\left( i\right) }$ is the gradient operator over the $i$ component of the
coordinate $\overrightarrow{r}$. The space dependence of $V_{ij}^{\left(
0\right) \left( l\right) }\left( \overrightarrow{R}\right) $ is valid for $%
\mid \overrightarrow{R}\mid \gg a_{s}$. The length $a_{s}$ is a
characteristic length of the long range spin-spin interation. The length $%
a_{s}$ is supposed to obey the inequality $\rho ^{-1/3}\gg a_{s}\gg a_{0}$.

In what follows, the Hamiltonian $\widehat{H}$ is written in terms of the
system of spin coherent states (SCS). The basis of the spin coherent states
(SCS) $\mid \overrightarrow{n}>$ for the spin of the paticle $S$ is used to
expand the creation and annihilation operators of particles $\widehat{\psi }%
^{+}$, $\widehat{\psi }$ instead of the states $\mid m>$. SCS $\mid
\overrightarrow{n}>$ is characterized by a unit vector $\overrightarrow{n}%
=\left( \sin \theta \cos \varphi ;\sin \theta \sin \varphi ;\cos \theta
\right) $, where $\theta $ and $\varphi $ are spherical angles \cite{Radcl},
\cite{Lieb}, \cite{7}. The SCS $\mid \overrightarrow{n}>$ can be obtained by
the action of the operator $\widehat{D}\left( \overrightarrow{n}\right) $ on
the eigenstate $\left\vert -S\right\rangle $ of the operator $\widehat{s}%
^{\left( z\right) }$ with the minimal eigenvalue $m=-S$
\begin{equation}
\mid \overrightarrow{n}>=\widehat{D}\left( \overrightarrow{n}\right) \mid
-S>,  \label{n}
\end{equation}

where
\begin{equation}
\widehat{D}\left( \overrightarrow{n}\right) =\exp \left( i\theta \left(
\overrightarrow{\varkappa }\cdot \widehat{\overrightarrow{s}}\right) \right)
.  \label{D}
\end{equation}%
Here $\overrightarrow{\varkappa }\mathbf{=}\left( \sin \varphi ,-\cos
\varphi ,0\right) $, where $\theta $ and $\varphi $ are spherical angles of
the vector $\overrightarrow{n}$. Thus, $\overrightarrow{\varkappa }\perp
\overrightarrow{n}$ and $\overrightarrow{\varkappa }\perp \overrightarrow{n}%
_{0}^{\left( c\right) }$\textbf{, }where $\overrightarrow{n}_{0}^{\left(
c\right) }=\left( 0;0;1\right) $. The spherical angle $\theta =0$ for the
vector $\overrightarrow{n}_{0}^{\left( c\right) }$, thus $\widehat{D}\left(
\overrightarrow{n}_{0}^{\left( c\right) }\right) =\widehat{1}$ , where $%
\widehat{1}$ is the unit operator,\ and, as the result, $\mid
\overrightarrow{n}_{0}^{\left( c\right) }>=\left\vert -S\right\rangle $. The
system of SCS forms the overcomplete system and its completeness relation is
\begin{equation}
\sum\limits_{\ \overrightarrow{n}}\mid \overrightarrow{n}><\overrightarrow{n}%
\mid =\widehat{1},  \label{1}
\end{equation}%
where $\sum\limits_{\overrightarrow{n}}$ means the integration over the
spherical angles $\theta $ and $\varphi $ and $\sum\limits_{\overrightarrow{n%
}}=\frac{\left( 2S+1\right) }{4\pi }\int \sin \theta d\theta d\varphi
\mathbf{.}$ The spin operator in the SCS representation can be represented
in the form

\begin{equation}
\widehat{\overrightarrow{s}}=-\left( S+1\right) \sum\limits_{\
\overrightarrow{n}}\mid \overrightarrow{n}>\overrightarrow{n}<%
\overrightarrow{n}\mid .  \label{s}
\end{equation}%
The transfer from the anihilation and creation operators $\widehat{\psi }%
_{m} $, $\widehat{\psi }_{m}^{+}$\ in $\mid m>$ state representation to SCS
representation $\widehat{\psi }_{\overrightarrow{n}}$ , $\widehat{\psi }_{%
\overrightarrow{n}}^{+}$ is defined by equations
\begin{equation}
\widehat{\psi }_{\overrightarrow{n}}=\sum\limits_{m=-S}^{S}<\overrightarrow{n%
}\mid m>\widehat{\psi }_{m}\text{, \ }\widehat{\psi }_{\overrightarrow{n}%
}^{+}=\sum\limits_{m=-S}^{S}\widehat{\psi }_{m}^{+}<m\mid \overrightarrow{n}>%
\text{, }  \label{5}
\end{equation}%
while the inverce transformation has the form

\begin{equation}
\widehat{\psi }_{m}=\sum\limits_{\overrightarrow{n}}<m\mid \overrightarrow{n}%
>\widehat{\psi }_{\overrightarrow{n}}\text{, \ \ \ }\widehat{\psi }%
_{m}^{+}=\sum\limits_{\overrightarrow{n}}\widehat{\psi }_{\overrightarrow{n}%
}^{+}<\overrightarrow{n}\mid m>  \label{5a}
\end{equation}

Note, that in the space of the functions being defined by Eq. (\ref{5}%
)~bra-ket product $<\overrightarrow{n}_{1}|\overrightarrow{n}_{2}>$ is
analogous to delta function $\delta \left( \overrightarrow{n}_{1}-%
\overrightarrow{n}_{2}\right) $.

The transition to SCS representation gives the Hamiltonians $\widehat{H}%
^{\left( 0\right) }$ and bare interaction Hamiltonians $\widehat{H}%
_{int}^{\left( 0\right) \left( sh\right) }$, $\widehat{H}_{int}^{\left(
0\right) \left( sp\mid sh\right) }$, $\widehat{H}_{int}^{\left( 0\right)
\left( l\right) }$ (\ref{H0}, \ref{H1}, \ref{H2}, \ref{Hlong}) in the form

\begin{equation*}
\widehat{H}^{\left( 0\right) }=\int d^{3}r\sum\limits_{\overrightarrow{n}%
_{1}}\widehat{\psi }_{\overrightarrow{n}_{1}}^{+}\left( t,\overrightarrow{r}%
\right) \left[ -\frac{1}{2m_{0}}\overrightarrow{\nabla }^{2}-\mu \right]
\widehat{\psi }_{\overrightarrow{n}_{1}}\left( t,\overrightarrow{r}\right)
\end{equation*}

\begin{equation*}
\widehat{H}_{int}^{\left( 0\right) \left( sh\right) }=g^{\left( 0\right)
}\int d^{3}r\sum\limits_{\overrightarrow{n}_{1},\overrightarrow{n}_{2}}%
\widehat{\psi }_{\overrightarrow{n}_{1}}^{+}\left( t,\overrightarrow{r}%
\right) \widehat{\psi }_{\overrightarrow{n}_{2}}\left( t,\overrightarrow{r}%
\right) \widehat{\psi }_{\overrightarrow{n}_{2}}^{+}\left( t,\overrightarrow{%
r}\right) \widehat{\psi }_{\overrightarrow{n}_{1}}\left( t,\overrightarrow{r}%
\right)
\end{equation*}

\begin{equation*}
\widehat{H}_{int}^{\left( 0\right) \left( sp\mid sh\right) }=\gamma
_{sh}^{\left( 0\right) }\left( S+1\right) ^{2}\sum\limits_{\overrightarrow{n}%
_{1},\overrightarrow{n}_{2}}\left( \overrightarrow{n}_{1}\cdot
\overrightarrow{n}_{2}\right) \int d^{3}r\widehat{\psi }_{\overrightarrow{n}%
_{1}}^{+}\left( t,\overrightarrow{r}\right) \widehat{\psi }_{\overrightarrow{%
n}_{2}}^{+}\left( t,\overrightarrow{r}\right) \widehat{\psi }_{%
\overrightarrow{n}_{2}}\left( t,\overrightarrow{r}\right) \widehat{\psi }_{%
\overrightarrow{n}_{1}}\left( t,\overrightarrow{r}\right)
\end{equation*}

\begin{eqnarray*}
\widehat{H}_{int}^{\left( 0\right) \left( l\right) } &=&\left( S+1\right)
^{2}\sum\limits_{\overrightarrow{n}_{1},\overrightarrow{n}_{2}}\int
d^{3}rd^{3}r^{\prime }\left( n_{1}^{\left( i\right) }n_{2}^{\left( j\right)
}\right) V_{i,j}^{\left( 0\right) \left( l\right) }\left( \overrightarrow{r},%
\overrightarrow{r}^{\prime }\right) \cdot \\
&&\cdot \widehat{\psi }_{\overrightarrow{n}_{1}}^{+}\left( t,\overrightarrow{%
r}\right) \widehat{\psi }_{\overrightarrow{n}_{2}}^{+}\left( t,%
\overrightarrow{r}^{\prime }\right) \widehat{\psi }_{\overrightarrow{n}%
_{2}}\left( t,\overrightarrow{r}^{\prime }\right) \widehat{\psi }_{%
\overrightarrow{n}_{1}}\left( t,\overrightarrow{r}\right)
\end{eqnarray*}

Further, a low-energy behavior of the system, determined by the excitations
with small momenta $\left\vert \overrightarrow{p}\right\vert \ll \sqrt{%
2m_{0}\mu }$, is considered. The contribution of virtual large momenta $%
\left\vert \overrightarrow{p}\right\vert \gg \sqrt{2m\mu }$ renormalize
essentially the bare interaction between low energy particles. In the case
of the low density Bose gas the renormalized interaction is defined by the
sum of the ladder diagrams. First, the ladder diagrams with small external
momenta involving only the short range interactions $\widehat{H}%
_{int}^{\left( 0\right) \left( sh\right) }$, $\widehat{H}_{int}^{\left(
0\right) \left( sp\mid sh\right) }$ are summarized. Due to the independence
of the momentum transfer of these interactions each link of the ladder
diagam has the divergence due to the integration over the internal momentum
corresponding to the internal bare Green functions $G^{\left( 0\right)
}\left( p\right) $, where $G^{\left( 0\right) }\left( p\right) =\left(
\omega +\mu -\frac{1}{2m_{0}}\overrightarrow{p}^{2}+i\delta \right) ^{-1}$.
This divergence is cut off by a momentum $\Lambda _{0}=1/a_{0}$, while the
cutoff momentum $\Lambda _{0}$ as well as $a_{0}$ are supposed independent
of the spins of interacting particles. The summation of the ladder diagrams
with the interactions $\widehat{H}_{int}^{\left( 0\right) \left( sh\right) }$
and $\widehat{H}_{int}^{\left( 0\right) \left( sp\mid sh\right) }$ gives the
renormalization of the short range interaction $\widehat{H}_{int}^{\left(
sh\right) }$ in the form

\begin{equation*}
\widehat{H}_{int}^{\left( sh\right) }=\int
d^{3}r\sum\limits_{i=1}^{3}\sum\limits_{\overrightarrow{n}_{1},%
\overrightarrow{n}_{2}}\Gamma _{\overrightarrow{n}_{1},\overrightarrow{n}%
_{2}}^{\left( sh\right) }\widehat{\psi }_{\overrightarrow{n}_{1}}^{+}\left(
t,\overrightarrow{r}\right) \widehat{\psi }_{\overrightarrow{n}%
_{2}}^{+}\left( t,\overrightarrow{r}\right) \widehat{\psi }_{\overrightarrow{%
n}_{2}}\left( t,\overrightarrow{r}\right) \widehat{\psi }_{\overrightarrow{n}%
_{1}}\left( t,\overrightarrow{r}\right)
\end{equation*}

where

\begin{equation*}
\Gamma _{\overrightarrow{n}_{1},\overrightarrow{n}_{2}}^{\left( sh\right) }=%
\frac{g^{\left( 0\right) }+\gamma _{sh}^{\left( 0\right) }\left( S+1\right)
^{2}\left( \overrightarrow{n}_{1}\cdot \overrightarrow{n}_{2}\right) }{1-%
\left[ g^{\left( 0\right) }+\gamma _{sh}^{\left( 0\right) }\left( S+1\right)
^{2}\left( \overrightarrow{n}_{1}\cdot \overrightarrow{n}_{2}\right) \right]
T_{0}}
\end{equation*}

and

\begin{equation*}
T_{0}=i\int \frac{d^{4}p_{1}}{\left( 2\pi \right) ^{4}}G^{\left( 0\right)
}\left( p_{1}\right) G^{\left( 0\right) }\left( -p_{1}\right) =-\frac{m_{0}}{%
2\pi ^{2}a_{0}}
\end{equation*}

In the present paper we assume that $g^{\left( 0\right) }+\gamma
_{sh}^{\left( 0\right) }\left( S+1\right) ^{2}\left( \overrightarrow{n}%
_{1}\cdot \overrightarrow{n}_{2}\right) $ is positive and suffisiently
large, so that $\left( g^{\left( 0\right) }+\gamma _{sh}^{\left( 0\right)
}\left( S+1\right) ^{2}\right) \mid T_{0}\mid >>1$. In this case the
interaction constant $\Gamma _{\overrightarrow{n}_{1},\overrightarrow{n}%
_{2}}^{\left( sh\right) }$ can be put independent of $\overrightarrow{n}_{1},%
\overrightarrow{n}_{2}$ and equal to

\begin{equation*}
\Gamma _{\overrightarrow{n}_{1},\overrightarrow{n}_{2}}^{\left( sh\right)
}=g=\frac{1}{\mid T_{0}\mid }
\end{equation*}

To find the total renormalized interaction the dagrams with the bare long
range interaction $\widehat{H}_{int}^{\left( 0\right) \left( l\right) }$,
besides the diagram's block of ladder diagrams with the interaction vertexes
$\widehat{H}_{int}^{\left( sh\right) }$, should be taken into account for
the summation of the total sequence of ladder diagrams. The smallness of the
coupling constant $\gamma _{l}^{\left( 0\right) }$ compared with the
coupling constant $g$ is supposed, so that $g>>\gamma _{l}^{\left( 0\right)
}S^{2}$ or $\frac{m_{0}\gamma _{l}^{\left( 0\right) }S^{2}}{2\pi ^{2}a_{s}}%
<<1$. In this case taking into account that $a_{0}/a_{s}<<1$ and $R>>a_{s}$
the renormalization of the long range spin-spin interaction can be
considered in the first order over $\widehat{H}_{int}^{\left( 0\right)
\left( l\right) }$. As a result, the renormalized long range spin-spin
interaction gives the attractive short range term and takes the form

\begin{eqnarray*}
\widehat{H}_{int}^{\left( l\right) } &=&\left( S+1\right) ^{2}\sum\limits_{%
\overrightarrow{n}_{1},\overrightarrow{n}_{2}}\int d^{3}rd^{3}r^{\prime
}\left( n_{1}^{\left( i\right) }n_{2}^{\left( j\right) }\right)
V_{i,j}^{\left( r\right) \left( l\right) }\left( \overrightarrow{R}\right)
\cdot \\
&&\cdot \widehat{\psi }_{\overrightarrow{n}_{1}}^{+}\left( t,\overrightarrow{%
r}\right) \widehat{\psi }_{\overrightarrow{n}_{2}}^{+}\left( t,%
\overrightarrow{r}^{\prime }\right) \widehat{\psi }_{\overrightarrow{n}%
_{2}}\left( t,\overrightarrow{r}^{\prime }\right) \widehat{\psi }_{%
\overrightarrow{n}_{1}}\left( t,\overrightarrow{r}\right)
\end{eqnarray*}

where
\begin{equation}
V_{ij}^{\left( r\right) \left( l\right) }\left( \overrightarrow{R}\right)
=-\gamma _{l}^{\left( 0\right) }\nabla ^{\left( i\right) }\nabla ^{\left(
j\right) }\frac{1}{\mid \overrightarrow{R}\mid }-\frac{8}{3}\frac{a_{0}}{%
a_{s}}\gamma _{l}^{\left( 0\right) }\delta \left( \overrightarrow{R}\right)
\delta _{i,j},  \label{renorm-ss}
\end{equation}

Note, that the zero momentum transfer Fourier transformation of the bare
interaction $V_{ij}^{\left( 0\right) \left( l\right) }\left( \overrightarrow{%
R}\right) $ vanishes, if one put $i=j$ and regularizes the interaction $%
V_{ij}^{\left( 0\right) \left( l\right) }\left( \overrightarrow{R}\right) $
by the factor $\exp \left( -\varkappa R\right) $ supposing that $\varkappa
\longrightarrow 0$. For this reason, the second term in Eq. (\ref{renorm-ss}%
) should be kept in spite of the fact that it is small over the parameter $%
a_{0}/a_{s}$.

Using (\ref{1}), (\ref{s}), (\ref{5}) one can determine the generating
functional $Z\left[ J,\overline{J}\right] $ in the SCS representation as

\begin{equation}
Z\left[ J,\overline{J}\right] =\int D\psi D\overline{\psi }\exp \left(
iS_{0}+iS_{int}^{\left( c\right) }+iS_{int}^{\left( s\right) }+iS_{J}\right)
,  \label{Z1}
\end{equation}%
where $S_{0}$ is a free action, $S_{int}^{\left( c\right) }$ and $%
S_{int}^{\left( s\right) }$ are the parts of the action determined by the
spin independent and spin dependent renormalized interactions, respectively,
the term $S_{J}$ is the part of the action connected with infinitesimal
sources $J_{\overrightarrow{n}}\left( t,\overrightarrow{r}\right) .$Here%
\begin{equation}
S_{0}=\int dtd^{3}r\sum\limits_{\ \mathbf{n}}\left\{ \overline{\psi }_{\
\overrightarrow{n}}\left( t,\overrightarrow{r}\right) \left( i\partial
_{t}+\mu +\frac{1}{2m}\widehat{\Delta }\right) \psi _{\ \overrightarrow{n}%
}\left( t,\overrightarrow{r}\right) \right\}  \label{S0}
\end{equation}%
\begin{equation}
S_{int}^{\left( \rho \right) }=-g\int dtd^{3}r\rho ^{2}\left( t,%
\overrightarrow{r}\right)  \label{SintC}
\end{equation}%
\begin{equation}
S_{int}^{\left( s\right) }=-\int dtd^{3}rd^{3}r^{\prime }\left\{ S^{\left(
i\right) }\left( t,\overrightarrow{r}\right) V_{ij}^{\left( r\right) \left(
l\right) }\left( \overrightarrow{r}-\ \overrightarrow{r}^{\prime }\right)
S^{\left( j\right) }\left( t,\overrightarrow{r}^{\prime }\right) \right\}
\label{SintS}
\end{equation}%
\begin{equation}
S_{J}=\int dtd^{3}r\sum\limits_{\ \overrightarrow{n}}\overline{J}_{%
\overrightarrow{n}\ }\left( t,\overrightarrow{r}\right) \psi _{%
\overrightarrow{n}}\left( t,\overrightarrow{r}\right) +J_{\overrightarrow{n}%
}\left( t,\overrightarrow{r}\right) \overline{\psi }_{\overrightarrow{n}%
}\left( t,\overrightarrow{r}\right)  \label{SJ}
\end{equation}%
The density of particles $\rho \left( t,\overrightarrow{r}\right) $ and the
spin density vector $\overrightarrow{S}\left( t,\overrightarrow{r}\right) $
in SCS representation have the form
\begin{equation}
\rho \left( t,\overrightarrow{r}\right) =\sum\limits_{\mathbf{n}}\overline{%
\psi }_{\overrightarrow{n}}\left( t,\overrightarrow{r}\right) \psi _{%
\overrightarrow{n}}\left( t,\overrightarrow{r}\right) ,  \label{ro}
\end{equation}%
\begin{equation}
\overrightarrow{S}\left( t,\overrightarrow{r}\right) =\sum\limits_{\mathbf{n}%
}\overline{\psi }_{\overrightarrow{n}}\left( t,\overrightarrow{r}\right)
\overrightarrow{n}\psi _{\overrightarrow{n}}\left( t,\overrightarrow{r}%
\right) .  \label{Sp}
\end{equation}

Using the equality
\begin{equation*}
-\int d^{3}rd^{3}r^{\prime }\sum\limits_{i,j=1}^{3}S^{\left( i\right)
}\left( \overrightarrow{r}\right) \left( \nabla ^{\left( i\right) }\nabla
^{\left( j\right) }\frac{1}{\mid \overrightarrow{R}\mid }\right) S^{\left(
j\right) }\left( \overrightarrow{r}^{\prime }\right) =
\end{equation*}
\begin{equation*}
=4\pi \int d^{3}r\left\{ \left( \mathbf{curl}\overrightarrow{S}\left(
\overrightarrow{r}\right) \cdot \widehat{\Delta }^{-1}\mathbf{curl}%
\overrightarrow{S}\left( \overrightarrow{r}\right) \right) +\left(
\overrightarrow{S}\left( \overrightarrow{r}\right) \cdot \overrightarrow{S}%
\left( \overrightarrow{r}\right) \right) \right\}
\end{equation*}

, where $\widehat{\Delta }^{-1}$ is the inverse Laplace operator, we can
write the spin dependent part of the action as the sum of two terms. The
first one is the long range part and the second one is the short range part.

\begin{equation*}
S_{int}^{\left( s\right) }=S_{int}^{\left( s1\right) }+S_{int}^{\left(
s2\right) }
\end{equation*}%
\
\begin{equation}
S_{int}^{\left( s1\right) }=-4\pi \gamma _{l}^{\left( 0\right) }\int
dtd^{3}r\left\{ \text{$\mathbf{curl}$}\overrightarrow{S}\left( t,%
\overrightarrow{r}\right) \widehat{\Delta }^{-1}\mathbf{curl}\overrightarrow{%
S}\left( t,\overrightarrow{r}\right) \right\}  \label{Sr1}
\end{equation}%
\begin{equation}
S_{int}^{\left( s2\right) }=-4\pi \gamma _{l}^{\left( 0\right) }\int
dtd^{3}r\left\{ \left( 1-\frac{2}{3}\frac{a_{0}}{a_{s}}\right) \left(
\overrightarrow{S}\left( t,\overrightarrow{r}\right) \cdot \overrightarrow{S}%
\left( t,\overrightarrow{r}\right) \right) \right\}  \label{Sr2}
\end{equation}

Using the Habbard-Stratanovich transformation, the long range interaction (%
\ref{Sr1}) can be represented as the interaction via the exchange by the
virtual magnetic field $\overrightarrow{H}=$curl$\overrightarrow{A}\mathbf{,}
$ where $\overrightarrow{A}$ is a vector potential obeying the gauge $%
\mathrm{div}\overrightarrow{A}=0$. Due to this transformation the generation
functional $Z[J,\overline{J}]$ takes the form
\begin{equation}
Z[J,\overline{J}]=\int \prod\limits_{\overrightarrow{n}}D\psi _{%
\overrightarrow{n}}D\overline{\psi }_{\overrightarrow{n}}D\overrightarrow{A}%
\exp \left\{ i\left[ S_{act}\left[ \psi _{\overrightarrow{n}};\overline{\psi
}_{\overrightarrow{n}};\overrightarrow{A}\right] +S_{J}\right] \right\}
\label{Z}
\end{equation}%
\qquad \qquad\

where

\begin{equation}
S_{act}=\int dtd^{3}r\left\{
\begin{array}{c}
\sum\limits_{\overrightarrow{n}}\left( \overline{\psi }_{\overrightarrow{n}%
}\left( G^{\left( 0\right) }\right) ^{-1}\psi _{\overrightarrow{n}}\right) -
\\
-\frac{1}{2}g\rho ^{2}-4\pi \gamma \left( 1-\frac{2}{3}\frac{a_{0}}{a_{s}}%
\right) \left( \overrightarrow{S}\cdot \overrightarrow{S}\right) + \\
+\sqrt{2\gamma }\left( \overrightarrow{A}\cdot \mathbf{curl}\overrightarrow{S%
}\right) +\frac{1}{8\pi }\left( \overrightarrow{A}\cdot \widehat{\Delta }%
\overrightarrow{A}\right)%
\end{array}%
\right\}  \label{Act}
\end{equation}

and

\begin{equation*}
\mathrm{div}\overrightarrow{A}=0\text{; \ \ \ \ \ }\overrightarrow{H}=%
\mathbf{curl}\overrightarrow{A}
\end{equation*}

\begin{equation*}
\left( G^{\left( 0\right) }\right) ^{-1}=i\partial _{t}+\mu +\frac{1}{2m_{0}}%
\widehat{\Delta }
\end{equation*}

Gross - Pitaevskii (GPE) equations can be obtained from the action (\ref{Act}%
) by equating to zero the first variation of this action with respect to the
fields $\psi _{\overrightarrow{n}}$, $\overline{\psi }_{\overrightarrow{n}}$%
, $\overrightarrow{A}$. These equations take the form

\begin{equation}
\left( \left( G_{0}\right) ^{-1}-g\rho \right) \psi _{\overrightarrow{n}%
}-\left( S+1\right) \sqrt{2\gamma }\ \sum\limits_{\overrightarrow{n}_{1}}<%
\overrightarrow{n}\mid \overrightarrow{n}_{1}>\left( \overrightarrow{n}%
_{1}\cdot \widetilde{\overrightarrow{H}}\right) \psi _{\overrightarrow{n}%
_{1}}=0  \label{GP1}
\end{equation}

\begin{equation}
\frac{1}{4\pi }\widehat{\Delta }\overrightarrow{A}+\sqrt{2\gamma }\mathrm{%
\mathbf{curl}}\overrightarrow{S}=0,  \label{GP2}
\end{equation}

where

\begin{equation*}
\widetilde{\overrightarrow{H}}=\overrightarrow{H}\mathbf{-}\frac{8\pi \left(
\gamma -\widetilde{\gamma }\right) }{\sqrt{2\gamma }}\overrightarrow{S}
\end{equation*}

and $\left( G_{0}\right) ^{-1}=i\partial _{t}+\mu +\frac{1}{2m_{0}}\widehat{%
\Delta }$;$~\widetilde{\gamma }=\frac{2}{3}\frac{a_{0}}{a_{s}}\gamma $. From
Eq. (\ref{GP2}) it can be easily obtained that the magnetic field $%
\overrightarrow{H}$ in the momentum representation takes the form
\begin{equation}
\overrightarrow{H}_{\overrightarrow{k}}=4\pi \sqrt{2\gamma }\left[
\overrightarrow{S}_{\overrightarrow{k}}-\overrightarrow{e}_{\overrightarrow{k%
}}\left( \overrightarrow{e}_{\overrightarrow{k}}\cdot \overrightarrow{S}_{%
\overrightarrow{k}}\right) \right] ,  \label{GP2a}
\end{equation}%
where $\overrightarrow{e}_{\overrightarrow{k}}=\overrightarrow{k}\mathbf{/}%
\left\vert \overrightarrow{k}\right\vert $. Using (\ref{GP2a}), one obtains
from (\ref{GP1}) GPE in the momentum representation%
\begin{gather}
\left( i\partial _{t}+\mu -\frac{1}{2m}\overrightarrow{k}^{2}-g\rho _{-%
\overrightarrow{k}}\right) \psi _{\overrightarrow{n};\overrightarrow{k}}-
\label{GP1a} \\
-8\pi \left( S+1\right) \sum\limits_{i=1}^{3}\sum\limits_{\overrightarrow{n}%
_{1}}<\overrightarrow{n}\mid \overrightarrow{n}_{1}>\overrightarrow{\Gamma }%
_{\overrightarrow{k}}^{\left( i\right) }\overrightarrow{S}_{-\overrightarrow{%
k}}^{\left( i\right) }\psi _{\overrightarrow{n}_{1};\overrightarrow{k}}=0.
\notag
\end{gather}

where $S_{-\overrightarrow{k}}^{\left( i\right) }$, $n_{1}^{\left( i\right)
} $, $e_{\overrightarrow{k}}^{\left( i\right) }$ are $x,y,z$ components of
the vectors $\overrightarrow{S}_{-\overrightarrow{k}}$, $\overrightarrow{n}%
_{1}$, $\overrightarrow{e}_{\overrightarrow{k}}$, and

\begin{equation}
\overrightarrow{\Gamma }_{\overrightarrow{k}}^{\left( i\right) }=\widetilde{%
\gamma }n_{1}^{\left( i\right) }-\gamma \left( \overrightarrow{n}_{1}\mathbf{%
\cdot }\overrightarrow{e}_{\overrightarrow{k}}\right) \overrightarrow{e}_{%
\overrightarrow{k}}^{\left( i\right) }  \label{Gamm}
\end{equation}

To find the condensate field $\psi _{\overrightarrow{n}}$ one should
transfer to the limit $\overrightarrow{k}\longrightarrow 0$ in Eqs. (\ref%
{GP1a}), (\ref{GP2a}). Yet, the limit of $\overrightarrow{e}_{%
\overrightarrow{k}}$ for $\overrightarrow{k}\longrightarrow 0$ is not
defined. Taking into account that $\left( \overrightarrow{k}\mathbf{\cdot }%
\overrightarrow{H}_{\overrightarrow{k}}\right) =0$ for arbitrary $%
\overrightarrow{k}$, the limit of $\overrightarrow{e}_{\overrightarrow{k}}$
for $\overrightarrow{k}\longrightarrow 0$ can be extended to the unit vector
$\overrightarrow{e}_{0}$ obeying the condition $\left( \overrightarrow{e}_{0}%
\mathbf{\cdot }\overrightarrow{H}_{0}\right) =0$, where $\overrightarrow{H}%
_{0}=\overrightarrow{H}_{\overrightarrow{k}=0}$\textbf{.} The condition $%
\left( \overrightarrow{e}_{\overrightarrow{k}}\mathbf{\cdot }\overrightarrow{%
H}_{\overrightarrow{k}}\right) =0$ is valid for $\overrightarrow{k}\neq $~$%
0. $ This condition is extended to $\overrightarrow{k}=0$ and the values $%
\overrightarrow{e}_{0}\mathbf{~}$and $\overrightarrow{H}_{0}$ are supposed
to obey a similar condition $\left( \overrightarrow{e}_{0}\mathbf{\cdot }%
\overrightarrow{H}_{0}\right) =0$. If $\overrightarrow{H}_{0}$ is nonzero $%
\overrightarrow{H}_{0}\neq 0$ the spin density $\overrightarrow{S}_{0}=$ $%
\overrightarrow{S}_{\overrightarrow{k}=0}$ is parallel to $\overrightarrow{H}%
_{0}$ and $\left( \overrightarrow{e}_{0}\mathbf{\cdot }\overrightarrow{S}%
_{0}\right) =0.$ Thus, in the case $\overrightarrow{k}=$~$0$\ Eq. (\ref{GP2a}%
) gives

\begin{equation}
\overrightarrow{H}_{0}=\ 4\pi \sqrt{2\gamma }\overrightarrow{S}_{0}
\label{HS}
\end{equation}

The field of BEC $\psi _{\overrightarrow{n}}^{\left( c\right) }$ is
spatially homogeneous and time independent. It obeys Eq. (\ref{GP1a}) for $%
\overrightarrow{k}=0$, and in SCS representation can be found in the form%
\begin{equation}
\psi _{\overrightarrow{n}}^{\left( c\right) }=\sqrt{\rho ^{c}}<%
\overrightarrow{n}\mid \overrightarrow{n}_{0}^{\left( c\right) }>  \label{BC}
\end{equation}%
here $\rho ^{\left( c\right) }$ is the density of BEC. Both the spin density
of BEC $\overrightarrow{S}^{\left( c\right) }$ and the effective magnetic
field of BEC $\overrightarrow{H}^{\left( c\right) }=\ 4\pi \sqrt{2\gamma }%
\overrightarrow{S}^{\left( c\right) }$ are spatially independent. $%
\overrightarrow{S}^{\left( c\right) }$ is obtained after substitution of Eq.
(\ref{BC}) into Eq. (\ref{Sp}) \
\begin{equation}
\overrightarrow{S}^{\left( c\right) }=-(S+1)\sum\limits_{\overrightarrow{n}}%
\overline{\psi }_{\overrightarrow{n}}^{\left( c\right) }\overrightarrow{n}%
\psi _{\overrightarrow{n}}^{\left( c\right) }=-S\rho ^{\left( c\right) }%
\overrightarrow{n}_{0}^{\left( c\right) }  \label{Sc}
\end{equation}

Substituting Eq. (\ref{Sc}) into (\ref{GP1a}) and taking into account that $%
\left( \overrightarrow{e}_{0}\mathbf{\cdot }\overrightarrow{H}_{0}\right)
=\left( \overrightarrow{e}_{0}\mathbf{\cdot }\overrightarrow{S}_{0}\right)
=0 $ one obtains
\begin{equation}
\left[ i\partial _{t}+\mu -\left( g-8\pi \widetilde{\gamma }S^{2}\right)
\rho ^{\left( c\right) }\right] \psi _{\overrightarrow{n}}=0,  \label{EqGPn}
\end{equation}%
Requiring time independence of the field of BEC $\psi _{\overrightarrow{n}%
}^{\left( c\right) }$, one obtains from Eq. (\ref{EqGPn}) the expression for
the chemical potential $\mu $
\begin{equation}
\mu =\left( g-8\pi \widetilde{\gamma }S^{2}\right) \rho ^{\left( c\right) }
\label{mu}
\end{equation}

The substituton of Eq.(\ref{Sc}) to Eq.(\ref{HS}) gives

\begin{equation*}
\overrightarrow{H}^{\left( c\right) }=4\pi \sqrt{2\gamma }\overrightarrow{S}%
^{\left( c\right) }
\end{equation*}

The equation of the spin density dynamics can be obtained from GPE after
three simple operations. The first one is the multiplication of Eq. (\ref%
{GP1}) by $\overrightarrow{n}\overline{\psi }_{\overrightarrow{n}}$ and
further summation of this equlity\ over $\overrightarrow{n}$. The second one
is the multplcation of the equation complex conjugate to Eq. (\ref{GP1}) by $%
\overrightarrow{n}\psi _{\overrightarrow{n}}$ and further summation\ over $%
\overrightarrow{n}$. The third one is the subtraction of these two obtained
equalities. As the result, the equations describing the dynamics of the spin
density is obtained
\begin{equation}
\partial _{t}S^{i}+\frac{1}{2m}\sum\limits_{j=1}^{3}\nabla ^{j}\Pi ^{ij}%
\mathbf{+}\sqrt{2\gamma }\left[ \widetilde{\overrightarrow{H}}\times
\overrightarrow{S}\right] ^{i}=0,  \label{LL}
\end{equation}%
here $\Pi ^{ij}$ is a spin flux tensor

\begin{equation}
\Pi ^{ij}=i\left( S+1\right) \sum\limits_{\mathbf{n}}\left[ \overline{\psi }%
_{\overrightarrow{n}}n^{i}\left( \nabla ^{j}\psi _{\overrightarrow{n}%
}\right) -\left( \nabla ^{j}\overline{\psi }_{\overrightarrow{n}}\right)
n^{i}\psi _{\overrightarrow{n}}\right] .  \label{Pain}
\end{equation}%
Note that in $\mid m>$ - representation the tensor $\Pi ^{ij}$ takes the form

\begin{equation}
\Pi ^{ij}=-i\sum\limits_{m_{1};m_{2}}\left[ \overline{\psi }_{m_{1}}\widehat{%
s}_{m_{1};m_{2}}^{i}\left( \nabla ^{j}\psi _{m_{2}}\right) -\left( \nabla
^{j}\overline{\psi }_{m_{1}}\right) \widehat{s}_{m_{1};m_{2}}^{i}\psi
_{m_{2}}\right]  \label{Paim}
\end{equation}

Eqs. (\ref{LL}) describing the ferromagnetic system of moving Bose particles
with nonzero spin are analogous to the Landau-Lifshits equations which
describe the spin dynamics of localised at lattice sites spins.

To obtain Green functions and spectra of the elementary exitations the
second order expansion of the action Eq.(\ref{Act}) over the field
fluctuations $\delta \psi _{\overrightarrow{n}}$, $\delta \overline{\psi }_{%
\overrightarrow{n}}$, $\delta \overrightarrow{A}$ is considered. The field
fluctuations are $\delta \psi _{\overrightarrow{n}}=\psi _{\overrightarrow{n}%
}-\psi _{\overrightarrow{n}}^{\left( c\right) }$, $\delta \overline{\psi }_{%
\overrightarrow{n}}=\overline{\psi }_{\overrightarrow{n}}-\overline{\psi }_{%
\overrightarrow{n}}^{\left( c\right) }$, $\delta \overrightarrow{A}=%
\overrightarrow{A}-\overrightarrow{A}^{\left( c\right) }$, where $\mathbf{%
curl}\overrightarrow{A}^{\left( c\right) }=\overrightarrow{H}^{\left(
c\right) }$,\ and the second order expansion of the action over them takes
the form of the functional $\delta S_{act}\left[ \delta \psi _{%
\overrightarrow{n}};\delta \overline{\psi }_{\overrightarrow{n}};\delta
\overrightarrow{A}\right] $ , where

\begin{equation}
\delta S_{act}=\int dtd^{3}r\left\{
\begin{array}{c}
\sum\limits_{\overrightarrow{n}}\delta \overline{\psi }_{\overrightarrow{n}%
}\left( G^{\left( 0\right) }\right) ^{-1}\delta \psi _{\overrightarrow{n}}-
\\
-g\delta \rho ^{\left( 2\right) }\rho ^{\left( c\right) }-\frac{1}{2}g\delta
\rho ^{\left( 1\right) }\delta \rho ^{\left( 1\right) }- \\
-8\pi \gamma \left( 1-\frac{2}{3}\frac{a_{0}}{a_{s}}\right) \left( \delta
\overrightarrow{S}^{\left( 2\right) }\cdot \overrightarrow{S}^{\left(
c\right) }\right) - \\
-4\pi \gamma \left( 1-\frac{2}{3}\frac{a_{0}}{a_{s}}\right) \left( \delta
\overrightarrow{S}^{\left( 1\right) }\cdot \delta \overrightarrow{S}^{\left(
1\right) }\right) + \\
+8\pi \gamma \left( \delta \overrightarrow{S}^{\left( 2\right) }\cdot
\overrightarrow{S}^{\left( c\right) }\right) + \\
+\sqrt{2\gamma }\left( \delta \overrightarrow{A}\cdot \mathbf{curl}\delta
\overrightarrow{S}^{\left( 1\right) }\right) \\
+\frac{1}{8\pi }\left( \delta \overrightarrow{A}\cdot \widehat{\Delta }\cdot
\delta \overrightarrow{A}\right)%
\end{array}%
\right\}  \label{DAct}
\end{equation}

and $\delta \rho ^{\left( 2\right) }=\sum\limits_{\overrightarrow{n}}\delta
\overline{\psi }_{\overrightarrow{n}}\delta \psi _{\overrightarrow{n}}$; $%
\delta \rho ^{\left( 1\right) }=\sum\limits_{\overrightarrow{n}}\left(
\delta \overline{\psi }_{\overrightarrow{n}}\psi _{\overrightarrow{n}%
}^{\left( c\right) }+\overline{\psi }_{\overrightarrow{n}}^{\left( c\right)
}\delta \psi _{\overrightarrow{n}}\right) $; $\delta \overrightarrow{S}%
^{\left( 2\right) }=-\left( S+1\right) \sum\limits_{\overrightarrow{n}%
}\delta \overline{\psi }_{\overrightarrow{n}}\overrightarrow{n}\delta \psi _{%
\overrightarrow{n}}$; $\delta S^{\left( 1\right) }=-\left( S+1\right)
\sum\limits_{\overrightarrow{n}}\left( \delta \overline{\psi }_{%
\overrightarrow{n}}\overrightarrow{n}\psi _{\overrightarrow{n}}^{\left(
c\right) }+\overline{\psi }_{\overrightarrow{n}}^{\left( c\right) }%
\overrightarrow{n}\delta \psi _{\overrightarrow{n}}\right) $. The generation
functional $Z$ of the fluctuations can be written as

\begin{equation}
Z\left[ J_{\overrightarrow{n}};\overline{J}_{\overrightarrow{n}}\right]
=\int \prod\limits_{\overrightarrow{n}}D\delta \overline{\psi }_{%
\overrightarrow{n}}D\delta \overline{\psi }_{\overrightarrow{n}}D\delta
\overrightarrow{A}\exp \left\{ i\delta S_{act}\left[ \delta \psi _{%
\overrightarrow{n}};\delta \overline{\psi }_{\overrightarrow{n}};\delta
\overrightarrow{A}\right] +\delta S_{J}\right\}  \label{DZ}
\end{equation}

where $\delta S_{J}=\int dtd^{3}r\sum\limits_{\overrightarrow{n}}\left(
\overline{J}_{\overrightarrow{n}}\delta \psi _{\overrightarrow{n}}+J_{%
\overrightarrow{n}}\delta \overline{\psi }_{\overrightarrow{n}}\right) $.
The integration of Eq.(\ref{DZ}) over $\delta \overrightarrow{A}$ gives

\begin{equation*}
\delta S_{act}\left[ \delta \psi ;\delta \overline{\psi }\right] =\int
dtd^{3}r\left\{
\begin{array}{c}
\sum\limits_{\overrightarrow{n}}\delta \overline{\psi }_{\overrightarrow{n}%
}\left( G^{\left( 0\right) }\right) ^{-1}\delta \psi _{\overrightarrow{n}%
}+8\pi \widetilde{\gamma }\left( \delta \overrightarrow{S}^{\left( 2\right)
}\cdot \overrightarrow{S}^{\left( c\right) }\right) - \\
-g\delta \rho ^{\left( 2\right) }\rho ^{\left( c\right) }+4\pi \widetilde{%
\gamma }\left( \delta \overrightarrow{S}^{\left( 1\right) }\cdot \delta
\overrightarrow{S}^{\left( 1\right) }\right) - \\
-\frac{1}{2}g\delta \rho ^{\left( 1\right) }\delta \rho ^{\left( 1\right)
}-4\pi \gamma \left( \delta \overrightarrow{S}^{\left( 1\right) }\cdot
\delta \overrightarrow{S}^{\left( 1\right) }\right) - \\
-4\pi \gamma \left( \mathbf{curl}\delta \overrightarrow{S}^{\left( 1\right)
}\cdot \widehat{\Delta }^{-1}\cdot \mathbf{curl}\delta \overrightarrow{S}%
^{\left( 1\right) }\right)%
\end{array}%
\right\}
\end{equation*}

Using the equality

$\int d^{3}r\left\{ \left( \mathbf{curl}\overrightarrow{S}\cdot \widehat{%
\Delta }^{-1}\cdot \mathbf{curl}\overrightarrow{S}\right) +\left(
\overrightarrow{S}\cdot \overrightarrow{S}\right) \right\} =\int \frac{d^{3}k%
}{\left( 2\pi \right) ^{3}}\left( \overrightarrow{e}_{\overrightarrow{k}%
}\cdot \overrightarrow{S}_{\overrightarrow{k}}\right) \left( \overrightarrow{%
e}_{\overrightarrow{k}}\cdot \overrightarrow{S}_{-\overrightarrow{k}}\right)
$

we can transform the action of fluctuations $\delta \psi _{\overrightarrow{n}%
}$, $\delta \overline{\psi }_{\overrightarrow{n}}$ to the form $\delta
S_{act}\left[ \delta \psi _{\overrightarrow{n}};\delta \overline{\psi }_{%
\overrightarrow{n}}\right] $

\begin{equation}
\delta S_{act}=\int \frac{dtd^{3}k}{\left( 2\pi \right) ^{3}}\left\{
\begin{array}{c}
\sum\limits_{\overrightarrow{n}}\delta \overline{\psi }_{\overrightarrow{n};%
\overrightarrow{k}}\left(
\begin{array}{c}
i\partial _{t}+\mu -\frac{1}{2m_{0}}\overrightarrow{k}^{2}-g\rho ^{\left(
c\right) } \\
+8\pi \widetilde{\gamma }S\left( S+1\right) \rho ^{\left( c\right) }\left(
\overrightarrow{n}\cdot \overrightarrow{n}_{0}\right)%
\end{array}%
\right) \delta \psi _{\overrightarrow{n};\overrightarrow{k}}- \\
-\left(
\begin{array}{c}
\frac{1}{2}g\delta \rho _{-\overrightarrow{k}}^{\left( 1\right) }\delta \rho
_{\overrightarrow{k}}^{\left( 1\right) }- \\
-4\pi \widetilde{\gamma }\left( \delta \overrightarrow{S}_{-\overrightarrow{k%
}}^{\left( 1\right) }\cdot \delta \overrightarrow{S}_{\overrightarrow{k}%
}^{\left( 1\right) }\right) + \\
+4\pi \gamma \left( \overrightarrow{e}_{\overrightarrow{k}}\cdot \delta
\overrightarrow{S}_{\overrightarrow{k}}^{\left( 1\right) }\right) \left(
\overrightarrow{e}_{\overrightarrow{k}}\cdot \delta \overrightarrow{S}_{-%
\overrightarrow{k}}^{\left( 1\right) }\right)%
\end{array}%
\right)%
\end{array}%
\right\}  \label{Sfluct}
\end{equation}

The fields $\psi _{\overrightarrow{n}}\left( t,\overrightarrow{r}\right) $, $%
\overline{\psi }_{\overrightarrow{n}}\left( t,\overrightarrow{r}\right) $
taking into account the small fluctuations $\delta \psi _{\overrightarrow{n}%
}\left( t,\overrightarrow{r}\right) $ , $\delta \overline{\psi }_{%
\overrightarrow{n}}\left( t,\overrightarrow{r}\right) $ near the stationary
and homogeneous condensate field $\psi _{\overrightarrow{n}}^{\left(
c\right) }$ can be represented in the form

\begin{eqnarray*}
\psi _{\overrightarrow{n}}\left( t,\overrightarrow{r}\right) &=&\Phi \left(
t,\overrightarrow{r}\right) <\overrightarrow{n}\mid \overrightarrow{n}%
_{0}\left( t,\overrightarrow{r}\right) > \\
\overline{\psi }_{\overrightarrow{n}}\left( t,\overrightarrow{r}\right) &=&%
\overline{\Phi }\left( t,\overrightarrow{r}\right) <\overrightarrow{n}%
_{0}\left( t,\overrightarrow{r}\right) \mid \overrightarrow{n}>
\end{eqnarray*}

The function $\Phi \left( t,\overrightarrow{r}\right) $ determines small
density-phase fluctuations of the field. It can be represented as

\begin{equation}
\Phi \left( t,\overrightarrow{r}\right) =\sqrt{\rho ^{\left( c\right) }}%
+\delta \Phi \left( t,\overrightarrow{r}\right)  \label{PsiUn}
\end{equation}

where $\delta \Phi \left( t,\overrightarrow{r}\right) $ is a small
fluctuation of $\Phi \left( t,\overrightarrow{r}\right) $. If there is no
fluctuations $\delta \Phi \left( t,\overrightarrow{r}\right) =0$ the
function $\Phi \left( t,\overrightarrow{r}\right) $ does not depend on $t$
and $\overrightarrow{r}$ and takes the form $\Phi \left( t,\overrightarrow{r}%
\right) =\sqrt{\rho ^{\left( c\right) }}$. The smallness of the fluctuations
$\delta \Phi \left( t,\overrightarrow{r}\right) $ means that $\mid \delta
\Phi \left( t,\overrightarrow{r}\right) \mid <<\sqrt{\rho ^{\left( c\right) }%
}$. The ket-vector $\mid \overrightarrow{n}_{0}\left( t,\overrightarrow{r}%
\right) >$ is a spin coherent state, and the function $\overrightarrow{n}%
_{0}\left( t,\overrightarrow{r}\right) $ is a field of unit vectors. Due to
the smallness of the fluctuations the field $\overrightarrow{n}_{0}\left( t,%
\overrightarrow{r}\right) $ can be represented as

\begin{equation}
\overrightarrow{n}_{0}\left( t,\overrightarrow{r}\right) =\overrightarrow{n}%
_{0}^{\left( c\right) }+\delta \overrightarrow{n}_{0}\left( t,%
\overrightarrow{r}\right)  \label{Dn0}
\end{equation}

The state $\mid \overrightarrow{n}_{0}\left( t,\overrightarrow{r}\right) >$
is the fluctuating one near the stationary and homogenious state of the
condensate $\mid \overrightarrow{n}_{0}^{\left( c\right) }>=\mid -S>$. \ The
vector $\delta \overrightarrow{n}_{0}\left( t,\overrightarrow{r}\right) $
determines the small fluctuations of the spin direction near the constant
unit vector $\overrightarrow{n}_{0}^{\left( c\right) }$ which determines the
direction of the condensate's spin density $\overrightarrow{S}^{\left(
c\right) }$. The smallness of the spin fluctuanions $\delta \overrightarrow{n%
}_{0}\left( t,\overrightarrow{r}\right) $ means that $\mid \delta
\overrightarrow{n}_{0}\left( t,\overrightarrow{r}\right) \mid <<1$, $\delta
\overrightarrow{n}_{0}\left( t,\overrightarrow{r}\right) \bot
\overrightarrow{n}_{0}^{\left( c\right) }$. The spin coherent state $\mid
\overrightarrow{n}_{0}\left( t,\overrightarrow{r}\right) >$ can be
represented in the form \cite{7}, \cite{Radcl}, \cite{Lieb}

\begin{equation}
\mid \overrightarrow{n}_{0}\left( t,\overrightarrow{r}\right) >=\widehat{D}%
\left( \overrightarrow{n}_{0}\left( t,\overrightarrow{r}\right) \right) \mid
\overrightarrow{n}_{0}^{\left( c\right) }>  \label{n0tr}
\end{equation}

The operator $\widehat{D}\left( \overrightarrow{n}_{0}\left( t,%
\overrightarrow{r}\right) \right) $ is $\widehat{D}\left( \overrightarrow{n}%
_{0}\left( t,\overrightarrow{r}\right) \right) =\exp \left( i\theta \left(
\overrightarrow{\varkappa }\cdot \widehat{\overrightarrow{s}}\right) \right)
$ (\ref{D}), where the vectors $\overrightarrow{\varkappa }$ and $%
\overrightarrow{n}_{0}$ can be represented via the spherical angles $\theta $
and $\varphi $ which depend on $t$ and $\overrightarrow{r}$, at that, $%
\overrightarrow{\varkappa }=\left( \sin \varphi ;-\cos \varphi ;0\right) $, $%
\overrightarrow{n}_{0}=\left( \sin \theta \cos \varphi ;\sin \theta \sin
\varphi ;\cos \theta \right) $, $\overrightarrow{n}_{0}^{\left( c\right)
}=\left( 0;0;1\right) $. From the form of $\overrightarrow{\varkappa }$ and $%
\overrightarrow{n}_{0}$ it is clear that $\overrightarrow{\varkappa }\perp
\overrightarrow{n}_{0}$ and $\overrightarrow{\varkappa }\perp
\overrightarrow{n}_{0}^{\left( c\right) }$. The smallness of the
fluctuations $\delta \overrightarrow{n}_{0}$ means the smallness of $\theta $%
. As the result, for the small fluctuations\ the vector of the spin
direction $\overrightarrow{n}_{0}$ takes the form $\overrightarrow{n}%
_{0}=\left( \theta \cos \varphi ;\theta \sin \varphi ;1\right) =\left(
\delta n_{0}^{\left( x\right) };\delta n_{0}^{\left( y\right) };1\right) $, $%
\theta \overrightarrow{\varkappa }=\left( \delta n_{0}^{\left( y\right)
};-\delta n_{0}^{\left( x\right) };0\right) $, here the terms proporsional
to $\theta ^{2}$ are neglected. Note that the fields $\delta n_{0}^{\left(
+\right) }=\delta n_{0}^{\left( x\right) }+i\delta n_{0}^{\left( y\right) }$%
, $\delta n_{0}^{\left( -\right) }=\delta n_{0}^{\left( x\right) }-i\delta
n_{0}^{\left( y\right) }$ can be writen as $\delta n_{0}^{\left( +\right)
}=\theta \cdot e^{i\varphi }$, $\delta n_{0}^{\left( -\right) }=\theta \cdot
e^{-i\varphi }$. Due to the smallness of $\theta $\ the operator $\widehat{D}%
\left( \overrightarrow{n}_{0}\right) $ can be rewritten as

\begin{equation}
\widehat{D}\left( \overrightarrow{n}_{0}\right) =\widehat{1}+\frac{1}{2}%
\left( \delta n_{0}^{\left( +\right) }\cdot \widehat{s}^{\left( -\right) }-%
\text{\ }\delta n_{0}^{\left( -\right) }\cdot \widehat{s}^{\left( +\right)
}\right)  \label{Ds}
\end{equation}

where $\widehat{s}^{\left( \pm \right) }=\widehat{s}^{\left( x\right) }\pm i%
\widehat{s}^{\left( y\right) }$. As a result, the state $\mid
\overrightarrow{n}_{0}>$ can be written via the spin fluctuations $\delta
\overrightarrow{n}_{0}$ as

\begin{equation}
\mid \overrightarrow{n}_{0}>=\widehat{D}\left( \overrightarrow{n}_{0}\right)
\mid \overrightarrow{n}_{0}^{\left( c\right) }>=\mid -S>-\sqrt{\frac{S}{2}}%
\delta n_{0}^{\left( -\right) }\mid -S+1>  \label{n0dn0}
\end{equation}

Due to Eq.(\ref{n0dn0}) the field fluctuations $\delta \psi _{%
\overrightarrow{n}}$ and $\delta \overline{\psi }_{\overrightarrow{n}}$\ are
written as

\begin{equation}
\delta \psi _{\overrightarrow{n}}=\delta \Phi <\overrightarrow{n}\mid -S>-%
\sqrt{\frac{\rho ^{\left( c\right) }S}{2}}\delta n_{0}^{\left( -\right) }<%
\overrightarrow{n}\mid -S+1>  \label{deltaps}
\end{equation}

\begin{equation}
\delta \overline{\psi }_{\overrightarrow{n}}=\delta \overline{\Phi }<-S\mid
\overrightarrow{n}>-\sqrt{\frac{\rho ^{\left( c\right) }S}{2}}\delta
n_{0}^{\left( +\right) }<-S+1\mid \overrightarrow{n}>  \label{deltaps1}
\end{equation}

These equations can be rewritten as

\begin{equation}
\delta \psi _{\overrightarrow{n}}=\delta \Phi <\overrightarrow{n}\mid
-S>-\eta <\overrightarrow{n}\mid -S+1>  \label{dps1}
\end{equation}

\begin{equation}
\delta \overline{\psi }_{\overrightarrow{n}}=\delta \overline{\Phi }<-S\mid
\overrightarrow{n}>-\overline{\eta }<-S+1\mid \overrightarrow{n}>
\label{dps2}
\end{equation}

where

\begin{equation}
\eta =\sqrt{\frac{S\rho ^{\left( c\right) }}{2}}\delta n_{0}^{\left(
-\right) };\text{ \ \ \ }\overline{\eta }=\sqrt{\frac{S\rho ^{\left(
c\right) }}{2}}\delta n_{0}^{\left( +\right) }  \label{eta}
\end{equation}

The substitution of Eqs.(\ref{dps1}), (\ref{dps2}) to Eq.(\ref{Sfluct})
transforms the action of the field fluctuatons $\delta S_{act}\left[ \delta
\psi _{\overrightarrow{n}};\delta \overline{\psi }_{\overrightarrow{n}}%
\right] $ to the form $\delta S_{act}\left[ \delta \Phi ;\delta \overline{%
\Phi };\eta ;\overline{\eta }\right] $

\begin{equation}
\delta S_{act}=\delta S_{act}^{\left( \Phi \right) }+\delta S_{act}^{\left(
\eta \right) }+\delta S_{act}^{\left( \Phi ,\eta \right) }  \label{dS}
\end{equation}

where

\begin{equation}
\delta S_{act}^{\left( \Phi \right) }=\int \frac{d\omega d^{3}k}{\left( 2\pi
\right) ^{4}}\left(
\begin{array}{c}
\delta \overline{\Phi }_{k}\left( \omega -\frac{1}{2m_{0}}\overrightarrow{k}%
^{2}-\sigma _{\rho }\right) \delta \Phi _{k}- \\
-\frac{1}{2}\sigma _{\rho }\left( \delta \Phi _{-k}\delta \Phi _{k}+\delta
\overline{\Phi }_{k}\delta \overline{\Phi }_{-k}\right)%
\end{array}%
\right)  \label{dSpsai}
\end{equation}

\begin{equation}
\delta S_{act}^{\left( \eta \right) }=\int \frac{d\omega d^{3}k}{\left( 2\pi
\right) ^{4}}\left(
\begin{array}{c}
\overline{\eta }_{k}\left( \omega -\frac{1}{2m_{0}}\overrightarrow{k}%
^{2}-\sigma _{s}e_{k}^{\left( +\right) }e_{k}^{\left( -\right) }\right) \eta
_{k}- \\
-\frac{1}{2}\sigma _{s}\left( e_{k}^{\left( +\right) }e_{k}^{\left( +\right)
}\eta _{-k}\eta _{k}+e_{k}^{\left( -\right) }e_{k}^{\left( -\right) }%
\overline{\eta }_{k}\overline{\eta }_{-k}\right)%
\end{array}%
\right)  \label{dSeta}
\end{equation}

\begin{equation}
\delta S_{act}^{\left( \psi ,\eta \right) }=-\sqrt{\frac{S}{2}}\sigma
_{s}\int \frac{d\omega d^{3}k}{\left( 2\pi \right) ^{4}}e_{k}^{\left(
z\right) }\left(
\begin{array}{c}
e_{k}^{\left( -\right) }\overline{\eta }_{-k}\delta \Phi _{-k}+e_{k}^{\left(
+\right) }\eta _{k}\delta \overline{\Phi }_{k}+ \\
+e_{k}^{\left( +\right) }\eta _{k}\delta \Phi _{-k}+e_{k}^{\left( -\right) }%
\overline{\eta }_{-k}\delta \overline{\Phi }_{k}%
\end{array}%
\right)  \label{dSpsaieta}
\end{equation}

\begin{equation}
\sigma _{\rho }=\left( g-8\pi \widetilde{\gamma }S^{2}+8\pi \gamma
S^{2}\left( e_{k}^{\left( z\right) }\right) ^{2}\right) \rho ^{\left(
c\right) }  \label{sigmro}
\end{equation}

\begin{equation}
\sigma _{s}=4\pi \gamma S\rho ^{\left( c\right) }  \label{sigms}
\end{equation}

\begin{equation}
e_{k}^{\left( \pm \right) }=e_{k}^{\left( x\right) }\pm e_{k}^{\left(
y\right) };\text{ \ }\overrightarrow{e}_{k}=\frac{\overrightarrow{k}}{\mid
\overrightarrow{k}\mid }  \label{ek}
\end{equation}

The action $\delta S_{act}\left[ \delta \Phi ;\delta \overline{\Phi };\eta ;%
\overline{\eta }\right] $ Eq.(\ref{dS}) can be rewritten in the matrix form

\begin{equation}
\delta S_{act}=\int \frac{d\omega d^{3}k}{\left( 2\pi \right) ^{4}}%
\sum\limits_{\alpha ,\beta =1}^{4}\overline{\chi }_{i}\left( k\right) \left(
\widehat{G}^{-1}\right) _{i,j}\chi _{j}\left( k\right)  \label{dS1}
\end{equation}

The components of the field $\chi _{j}\left( k\right) $ are defined as $\chi
_{1}\left( k\right) =\delta \Phi _{k}$, $\chi _{2}\left( k\right) =\delta
\overline{\Phi }_{-k}$, $\chi _{3}\left( k\right) =\eta _{k}$, $\chi
_{4}\left( k\right) =\overline{\eta }_{-k}$, and $\overline{\chi }_{1}\left(
k\right) =\delta \overline{\Phi }_{k}$, $\chi _{2}\left( k\right) =\delta
\Phi _{-k}$, $\chi _{3}\left( k\right) =\overline{\eta }_{k}$, $\chi
_{4}\left( k\right) =\eta _{-k}$. The matrix $\widehat{G}^{-1}$ is the
inverse Green function of the exitations

\begin{equation}
\widehat{G}^{-1}=\left(
\begin{array}{cccc}
f_{\Phi } & -\sigma _{\rho } & -\sigma _{\Phi \eta }^{\left( +\right) } &
-\sigma _{\Phi \eta }^{\left( -\right) } \\
-\sigma _{\rho } & \widetilde{f}_{\Phi } & -\sigma _{\Phi \eta }^{\left(
+\right) } & -\sigma _{\psi \eta }^{\left( -\right) } \\
-\sigma _{\Phi \eta }^{\left( -\right) } & -\sigma _{\Phi \eta }^{\left(
-\right) } & f_{\eta } & -\sigma _{s}^{\left( --\right) } \\
-\sigma _{\Phi \eta }^{\left( +\right) } & -\sigma _{\Phi \eta }^{\left(
+\right) } & -\sigma _{s}^{\left( ++\right) } & \widetilde{f}_{\eta }%
\end{array}%
\right)  \label{Gm1}
\end{equation}

where the following denotations have the form $\sigma _{s}^{\left( +-\right)
}=\sigma _{s}e_{k}^{\left( +\right) }e_{k}^{\left( -\right) }$, $\sigma
_{s}^{\left( --\right) }=\sigma _{s}e_{k}^{\left( -\right) }e_{k}^{\left(
-\right) }$, $\sigma _{s}^{\left( ++\right) }=\sigma _{s}e_{k}^{\left(
+\right) }e_{k}^{\left( +\right) }$, $\sigma _{\Phi \eta }^{\left( \pm
\right) }=\sqrt{\frac{S}{2}}\sigma _{s}e_{k}^{\left( z\right) }e_{k}^{\left(
\pm \right) }$, and

\begin{eqnarray*}
f_{\Phi } &=&\omega -\frac{1}{2m_{0}}\overrightarrow{k}^{2}-\sigma _{\rho
}=f_{\psi }^{\left( 0\right) }-\sigma _{\rho } \\
f_{\eta } &=&\omega -\frac{1}{2m_{0}}\overrightarrow{k}^{2}-\sigma
_{s}e_{k}^{\left( +\right) }e_{k}^{\left( -\right) }=f_{\eta }^{\left(
0\right) }-\sigma _{s}e_{k}^{\left( +\right) }e_{k}^{\left( -\right) } \\
\widetilde{f}_{\Phi }\left( \omega \right) &=&f_{\Phi }\left( -\omega
\right) ;\text{ \ }\widetilde{f}_{\eta }\left( \omega \right) =f_{\eta
}\left( -\omega \right)
\end{eqnarray*}

The spectrum of the elementary exitations can be found from the equality to
zero of the determinant of the matrix $\widehat{G}^{-1}$. The calculation of
this determinant gives

\begin{equation*}
\det \widehat{G}^{-1}=\left\{
\begin{array}{c}
\left[ f_{\Phi }^{\left( 0\right) }\widetilde{f}_{\Phi }^{\left( 0\right)
}-\sigma _{\rho }\left( f_{\Phi }^{\left( 0\right) }+\widetilde{f}_{\Phi
}^{\left( 0\right) }\right) \right] \left[ f_{\eta }\widetilde{f}_{\eta
}-\sigma _{s}^{2}\left( e_{k}^{\left( +\right) }e_{k}^{\left( -\right)
}\right) ^{2}\right] - \\
-\frac{1}{2}\left( \sigma _{s}e_{k}^{\left( z\right) }\right)
^{2}Se_{k}^{\left( +\right) }e_{k}^{\left( -\right) }\left( f_{\Phi
}^{\left( 0\right) }+\widetilde{f}_{\Phi }^{\left( 0\right) }\right) \left(
f_{\eta }^{\left( 0\right) }+\widetilde{f}_{\eta }^{\left( 0\right) }\right)%
\end{array}%
\right\}
\end{equation*}

or

$\det \widehat{G}^{-1}=\left[ \omega ^{2}-\left( \frac{\overrightarrow{k}^{2}%
}{2m_{0}}\right) ^{2}-2\sigma _{\rho }\frac{\overrightarrow{k}^{2}}{2m_{0}}%
\right] \left[ \omega ^{2}-\left( \frac{\overrightarrow{k}^{2}}{2m_{0}}%
+\sigma _{s}\left( e_{k}^{\left( +\right) }e_{k}^{\left( -\right) }\right)
\right) ^{2}+\sigma _{s}^{2}\left( e_{k}^{\left( +\right) }e_{k}^{\left(
-\right) }\right) ^{2}\right] -$

$-2\sigma _{s}^{2}S\left( e_{k}^{\left( z\right) }\right) ^{2}\left(
e_{k}^{\left( +\right) }e_{k}^{\left( -\right) }\right) $

Thus, the spectrum of the elementary extations $\varepsilon _{%
\overrightarrow{k}}$ is defined by the equality

$\varepsilon _{\overrightarrow{k}}^{2}=\left( \frac{\overrightarrow{k}^{2}}{%
2m_{0}}\right) \left\{
\begin{array}{c}
\left( \sigma _{\rho }+\sigma _{s}e_{k}^{\left( +\right) }e_{k}^{\left(
-\right) }+\frac{\overrightarrow{k}^{2}}{2m_{0}}\right) \pm \\
\pm \sqrt{\left\{
\begin{array}{c}
\left( \sigma _{\rho }+\sigma _{s}e_{k}^{\left( +\right) }e_{k}^{\left(
-\right) }+\frac{\overrightarrow{k}^{2}}{2m_{0}}\right) ^{2}- \\
-\left( 2\sigma _{\rho }+\frac{\overrightarrow{k}^{2}}{2m_{0}}\right) \left(
\frac{\overrightarrow{k}^{2}}{2m_{0}}+2\sigma _{s}e_{k}^{\left( +\right)
}e_{k}^{\left( -\right) }\right) + \\
+2\left( \sigma _{s}e_{k}^{\left( z\right) }\right) ^{2}Se_{k}^{\left(
+\right) }e_{k}^{\left( -\right) }%
\end{array}%
\right\} }%
\end{array}%
\right\} $

In the case $e_{k}^{\left( +\right) }=e_{k}^{\left( -\right) }=0$, $%
e_{k}^{\left( z\right) }=1$ the spectrum has two independent branches. For
small momenta $\overrightarrow{k}^{2}/2m_{0}<<\sigma _{\rho }$ the first
branch is the sound like oscillations of the condensate module-phase and has
the form

\begin{equation}
\varepsilon _{\overrightarrow{k}}^{\left( 1\right) }=u_{\rho }\mid
\overrightarrow{k}\mid  \label{eps1}
\end{equation}

where $u_{\rho }=\sqrt{\frac{\sigma _{\rho }}{m_{0}}}$. The second branch is
the oscillations of the spin direction of the condensate and has the form

\begin{equation}
\varepsilon _{\overrightarrow{k}}^{\left( 2\right) }=\frac{1}{2m_{0}}%
\overrightarrow{k}^{2}  \label{eps2}
\end{equation}

In the case of nonzero $e_{k}^{\left( +\right) },$ $e_{k}^{\left( -\right) }$
and small momenta $\overrightarrow{k}^{2}/2m_{0}<<\sigma _{\rho }$; $%
\overrightarrow{k}^{2}/2m_{0}<<\sigma _{s}e_{k}^{\left( +\right)
}e_{k}^{\left( -\right) }$ the spectrum has two branches of the sound type

\begin{equation}
\varepsilon _{\overrightarrow{k}}^{\left( 1,2\right) }=\mid \overrightarrow{k%
}\mid \sqrt{\frac{\left( \sigma _{\rho }+\sigma _{s}e_{k}^{\left( +\right)
}e_{k}^{\left( -\right) }\right) \pm \sqrt{\left( \sigma _{\rho }-\sigma
_{s}e_{k}^{\left( +\right) }e_{k}^{\left( -\right) }\right) ^{2}+2\left(
\sigma _{s}e_{k}^{\left( z\right) }\right) ^{2}Se_{k}^{\left( +\right)
}e_{k}^{\left( -\right) }}}{2m_{0}}}  \label{eps12}
\end{equation}

Due to smallness of $\gamma $, so that $\gamma S^{2}<<g$, these two branches
of the spectrum take the form

\begin{equation}
\varepsilon _{\overrightarrow{k}}^{\left( 1,2\right) }=u^{\left( 1,2\right)
}\mid \overrightarrow{k}\mid  \label{e12}
\end{equation}

where $u^{\left( 1\right) }=\sqrt{\frac{\sigma _{\rho }}{m_{0}}}$, $%
u^{\left( 2\right) }=\sqrt{\frac{\sigma _{s}e_{k}^{\left( +\right)
}e_{k}^{\left( -\right) }}{m_{0}}}$.

In the case of nonzero $e_{k}^{\left( +\right) },$ $e_{k}^{\left( -\right) }$
and small momenta, such that $\sigma _{s}e_{k}^{\left( +\right)
}e_{k}^{\left( -\right) }<<\frac{1}{2m_{0}}\overrightarrow{k}^{2}<<\sigma
_{\rho }$, the spectrum has the same two branches as in the case for $%
e_{k}^{\left( +\right) }=e_{k}^{\left( -\right) }=0$, $e_{k}^{\left(
z\right) }=1$, i.e., $\varepsilon _{\overrightarrow{k}}^{\left( 1\right)
}=u_{\rho }\mid \overrightarrow{k}\mid $, $\varepsilon _{\overrightarrow{k}%
}^{\left( 2\right) }=\frac{1}{2m_{0}}\overrightarrow{k}^{2}$.

To find the Green function of the elementary exitations $\widehat{G}$ the
inverce to $\widehat{G}^{-1}$ matrix should be found. The elements of this
matrix have the form

\begin{equation}
G_{ij}=\left( -1\right) ^{i+j}\frac{\det \left( \widehat{A}_{ij}\right) }{%
\det \left( \widehat{G}^{-1}\right) }  \label{G}
\end{equation}

where $\widehat{A}_{ij}$ is the coresponding minor of the matrix $\widehat{G}%
^{-1}$, and $1\leq i,j\leq 4$.

In the case $e_{k}^{\left( +\right) }=e_{k}^{\left( -\right) }=0$, $%
e_{k}^{\left( z\right) }=1$ and small momenta $\mid \overrightarrow{k}\mid
^{2}<<\mu m_{0}$ the nonzero Green functions $G_{11}=G_{\overline{\Phi }\Phi
}$, $G_{21}=G_{\overline{\Phi }\overline{\Phi }}$, $G_{33}=G_{\overline{\eta
}\eta }$, $G_{34}=G_{\overline{\eta }\overline{\eta }}$ take the form
\begin{eqnarray}
G_{\overline{\Phi }\Phi } &=&\frac{\sigma _{\rho }}{\omega ^{2}-u_{\rho
}^{2}\mid \overrightarrow{k}\mid ^{2}+i\delta }  \label{G11} \\
G_{\overline{\Phi }\overline{\Phi }} &=&\frac{-\sigma _{\rho }}{\omega
^{2}-u_{\rho }^{2}\mid \overrightarrow{k}\mid ^{2}+i\delta }  \notag
\end{eqnarray}

\begin{eqnarray}
G_{\overline{\eta }\eta } &=&\frac{1}{\omega -\frac{1}{2m_{0}}\mid
\overrightarrow{k}\mid ^{2}+i\delta }  \label{G13} \\
G_{\overline{\eta }\overline{\eta }} &=&\frac{1}{-\omega -\frac{1}{2m_{0}}%
\mid \overrightarrow{k}\mid ^{2}+i\delta }  \notag
\end{eqnarray}

In the case $e_{k}^{\left( +\right) }\neq 0$, $e_{k}^{\left( -\right) }\neq
0 $ and small momenta $\frac{1}{2m_{0}}\overrightarrow{k}^{2}<<\sigma _{\rho
}$; $\frac{1}{2m_{0}}\overrightarrow{k}^{2}<<\sigma _{s}e_{k}^{\left(
+\right) }e_{k}^{\left( -\right) }$ these Green functions take the form

\begin{eqnarray}
G_{\overline{\Phi }\Phi } &=&\frac{\sigma _{\rho }\left( \omega ^{2}-\left(
u^{\left( 2\right) }\right) ^{2}\overrightarrow{k}^{2}\right) +S\sigma
_{s}^{2}\left( e_{k}^{\left( z\right) }\right) ^{2}\left( e_{k}^{\left(
+\right) }e_{k}^{\left( -\right) }\right) \frac{\overrightarrow{k}^{2}}{%
2m_{0}}}{\left( \omega ^{2}-\left( u^{\left( 1\right) }\right) ^{2}%
\overrightarrow{k}^{2}+i\delta \right) \left( \omega ^{2}-\left( u^{\left(
2\right) }\right) ^{2}\overrightarrow{k}^{2}+i\delta \right) }  \label{G21}
\\
G_{\overline{\Phi }\overline{\Phi }} &=&\frac{\sigma _{\rho }\left( \omega
^{2}-\left( u^{\left( 2\right) }\right) ^{2}\overrightarrow{k}^{2}\right)
+S\sigma _{s}^{2}\left( e_{k}^{\left( z\right) }\right) ^{2}e_{k}^{\left(
+\right) }e_{k}^{\left( -\right) }\frac{\overrightarrow{k}^{2}}{2m_{0}}}{%
\left( \omega ^{2}-\left( u^{\left( 1\right) }\right) ^{2}\overrightarrow{k}%
^{2}+i\delta \right) \left( \omega ^{2}-\left( u^{\left( 2\right) }\right)
^{2}\overrightarrow{k}^{2}+i\delta \right) }  \notag
\end{eqnarray}

\begin{eqnarray}
G_{\overline{\eta }\eta } &=&\sigma _{s}\left( e_{k}^{\left( +\right)
}e_{k}^{\left( -\right) }\right) \frac{\left( \omega ^{2}-\left( u^{\left(
1\right) }\right) ^{2}\overrightarrow{k}^{2}\right) +S\sigma _{s}\left(
e_{k}^{\left( z\right) }\right) ^{2}\left( \frac{\overrightarrow{k}^{2}}{%
2m_{0}}\right) }{\left( \omega ^{2}-\left( u^{\left( 1\right) }\right) ^{2}%
\overrightarrow{k}^{2}+i\delta \right) \left( \omega ^{2}-\left( u^{\left(
2\right) }\right) ^{2}\overrightarrow{k}^{2}+i\delta \right) }  \label{G23}
\\
G_{\overline{\eta }\overline{\eta }} &=&\sigma _{s}\left( e_{k}^{\left(
+\right) }e_{k}^{\left( +\right) }\right) \frac{\left( \omega ^{2}-\left(
u^{\left( 1\right) }\right) ^{2}\overrightarrow{k}^{2}\right) +S\sigma
_{s}\left( e_{k}^{\left( z\right) }\right) ^{2}\left( \frac{\overrightarrow{k%
}^{2}}{2m_{0}}\right) }{\left( \omega ^{2}-\left( u^{\left( 1\right)
}\right) ^{2}\overrightarrow{k}^{2}+i\delta \right) \left( \omega
^{2}-\left( u^{\left( 2\right) }\right) ^{2}\overrightarrow{k}^{2}+i\delta
\right) }  \notag
\end{eqnarray}

For $e_{k}^{\left( +\right) }\neq 0$, $e_{k}^{\left( -\right) }\neq 0$ and
small momenta, such that $\sigma _{s}e_{k}^{\left( +\right) }e_{k}^{\left(
-\right) }<<\frac{1}{2m_{0}}\overrightarrow{k}^{2}<<\sigma _{\rho }$, the
Green functions has the same form as in the case for $e_{k}^{\left( +\right)
}=e_{k}^{\left( -\right) }=0$, $e_{k}^{\left( z\right) }=1$. Note that if $%
e_{k}^{\left( +\right) }\neq 0$, $e_{k}^{\left( -\right) }\neq 0$\ there are
nonzero components of $\widehat{G}$ which define the correlations between
module-phase exititions and spin direction exitations. They are $G_{\Phi
\overline{\eta }}$ or $G_{\Phi \eta }$, for example.

\bigskip

\bigskip

\end{document}